# JALE: A new Python toolbox for coordinate-based neuroimaging meta-analysis using activation likelihood estimation

**Edna C. Cieslik** (ORCID: 0000-0001-5491-5837)[1,2,*], **Lennart Frahm**[1,3], **Felix Hoffstaedter**[1,2], **Robert Langner** (ORCID: 0000-0002-3237-001X)[1,2], **Changyi Kuang** (ORCID: 0000-0001-7033-7467) [1,2], **Peter T. Fox** (ORCID: 0000-0002-0465-2028)[4], **Federico Raimondo** (ORCID 0000-0003-4087-8259)[1,2], **Simon B. Eickhoff**[1,2], **Veronika I. Müller** (ORCID: 0000-0003-4382-3181)[1,2,*]

[1] Institute of Neuroscience und Medicine, INM-7, Research Centre Jülich, Germany;

[2] Institute of Systems Neuroscience, Medical Faculty and University Hospital Düsseldorf, Heinrich Heine University, Düsseldorf, Germany;

[3] Department of Psychiatry, Psychotherapy and Psychosomatics, School of Medicine, RWTH Aachen University, Aachen, Germany;

[4] Research Imaging Institute, University of Texas Health Science Center, San Antonio, USA; Departments of Radiology, Neurology, Psychiatry and Behavioral Sciences, and Physiology, University of Texas Health Science Center, San Antonio, USA

*corresponding author

Correspondence concerning this article should be addressed to Edna C. Cieslik, Institute of Neuroscience and Medicine, INM-7, Research Centre Jülich, Germany, email: e.cieslik@fz-juelich.de or Veronika I. Müller, Institute of Neuroscience and Medicine, INM-7, Research Centre Jülich, Germany, email: v.mueller@fz-juelich.de

**Abstract**

Coordinate-based neuroimaging meta-analysis using activation likelihood estimation (ALE) is commonly used to synthesize reported results of functional or structural magnetic resonance imaging studies. We here present JALE, a new, open-source Python package for conducting ALE meta-analyses. Besides computing standard ALE main effects and contrast analyses, this toolbox implements several novel analytical approaches including new options for thresholding (threshold-free cluster enhancement statistics, significance threshold prediction) as well as sensitivity and cluster analyses that were only available as in-house scripts in the past. The purpose of this paper is to provide an integrated conceptual foundation of recent developments in the ALE meta-analysis framework, with a particular focus on novel methodological implementations that can be executed in a straightforward and standardized manner using the new JALE toolbox.

**Keywords:** activation likelihood estimation, neuroimaging, fMRI, meta-analysis, toolbox

## 1. Introduction

Since its invention in the late 1980s, functional magnetic resonance imaging (fMRI) has rapidly evolved as a widely used technique to study brain–behaviour relationships in the healthy human brain as well as in psychiatric or neurological disorders. While the advantages of fMRI as a non-invasive, non-radioactive, and radiation-free technology for in-vivo neuroimaging have certainly contributed to its popularity, there are some constraints that limit the generalizability of fMRI-derived study results. In particular, study-specific effects and characteristics, such as typically low sample sizes as well as analytical and experimental flexibility, undermine the generalizability and replicability of individual fMRI studies (Botvinik-Nezer et al., 2020; Button et al., 2013; Carp, 2012; Turner et al., 2018). Additionally, the growing number fMRI studies being published every year poses an additional challenge in maintaining an overview of the multitude and heterogeneity of results and to distinguish spurious from replicable findings. One solution to these issues is using neuroimaging meta-analyses to quantitatively consolidate effects across individual studies.

The earliest meta-analysis tool developed for typically reported results in the neuroimaging literature (i.e., peak activation coordinates) was activation likelihood estimation (ALE), which soon became one of the most widely used approaches for coordinate-based neuroimaging meta-analysis. Since performing ALE meta-analyses became easily possible through the openly available software tool GingerALE (https://brainmap.org/ale/) in 2003, the number of published meta-analyses exploded. However, over time it became clear that even highly integrative meta-analyses may suffer from limited replicability due to untransparent decision making and reporting (Müller et al., 2017). This then led to the formulation of best-practice recommendations for various neuroimaging meta-analytic approaches in fundamental as well as clinical research (Müller et al., 2018; Tahmasian et al., 2019), which have recently been supplemented by a quality criteria checklist (Radua et al., 2026).

Since its introduction (Laird et al., 2005; Turkeltaub et al., 2002), the ALE approach has been consistently refined on the technical level as well as on the application side. ALE estimates the likelihood of overlap between the activation foci from different studies by accounting for the spatial uncertainty of effects through modeling a Gaussian kernel around each individual activation focus. While in its first implementation (Turkeltaub et al., 2002) the size of the spatial uncertainty had to be specified by the user, Eickhoff et al. (2009) provided empirically derived estimates of between-subject as well as between-template variability. Further, the revised algorithm assessed above-chance clustering across experiments, rather than foci, allowing random-effects inference and, hence, generalization of the results to the entire population of studies from which the analyzed sample was drawn (cf. Eickhoff et al., 2009). In 2012, Eickhoff et al. introduced a new analytical solution to compute the null distribution of ALE scores under the assumption of spatial independence and also complemented the correction procedure of controlling the false-discovery rate (FDR) with new approaches for correcting the voxel-level (vFWE) as well as cluster-level (cFWE) family-wise error rate. These thresholding and multiple-comparison correction procedures have been evaluated via extensive simulations (Eickhoff et al., 2016; Frahm et al., 2022; Frahm et al., 2023) a few years later.

Across the last two decades, Python (https://www.python.org) has become one of the standard programming languages in many scientific areas as it combines the advantages of an open-source software with ease of use and powerful computational capabilities. Further, its modular framework hosts thousands of third-party modules ("packages") supporting tasks from data structuring to analysis and visualisation. To make ALE and especially our new developments and implementations available for every researcher independent of license regulations, we have now translated our in-house MATLAB scripts into a Python-based toolbox called JALE (https://github.com/inm7/jale), This toolbox is not intended as a substitute to existing packages such as GingerALE or NiMARE (https://nbclab.github.io/nimare-paper; Salo et al., 2023). Rather, our intention was to provide the research community with a consolidated implementation of recently developed and evaluated analysis options (Frahm et al., 2022; Frahm et al., 2025) as well as with novel approaches that have been used as in-house tools but have not been publicly available before. Within the last five years, our group has developed, evaluated and implemented new approaches and applications for ALE meta-analysis including thresholding based on TFCE, correction for multiple comparisons based on machine-learning trained models, analyses accounting for potential biases due to sample size, robustness evaluation as well as cluster-analysis. So far, these new implementations have only been available within our group as either MATLAB or Python scrips and have been shared with individual researchers on request. Here we aim to make the new implementations accessible to the entire neuroimaging community in an open-source, easy-to-use and transparent format. The toolbox allows anyone running ALE meta-analyses out of the box, using recommended settings, while it also entails options to modify certain parameters, which may be useful for advanced users.

## 2. General description

JALE is openly available via github (https://github.com/inm7/jale), licensed under the MIT license, and is currently running under Python 3.11. It also contains an online Wiki, with detailed descriptions of how to set up Python on Mac OS as well as Linux, how to install JALE as well as how to set up a meta-analytic project folder. Once the user has set up Python and created a virtual environment, the single command "pip install JALE" will install JALE together with all necessary other Python packages. Further, the Wiki also includes an example project folder with all files that are necessary to run a test meta-analysis.

In general, the following three files (Fig. 1) are needed to run a meta-analysis with JALE:

a) an **experiment info file** (excel or csv format) containing the data to be meta-analyzed (Fig. 1A), which must include for each experiment: (i) experiment name, (ii) number of participants, (iii) x,y,z-coordinates, and (iv) coordinate space (MNI/Talairach) in this specific order. Additional project-specific tags can be included to run tag-specific meta-analyses.

b) an **analysis info file** (excel or csv format) defining which specific meta-analysis should be calculated and which experiments from the experiment info file should be included (Fig. 1B). The current algorithm allows performing standard main-effect meta-analyses, iterative probabilistic ALE (IPA), standard meta-analytic contrasts, balanced contrast analysis as well as meta-analytic clustering (see below for details).

c) a **config file** that defines the names of the respective experiment-info and analysis-info files as well as specifications of the ALE parameters (Fig. 1C). In general, this file entails the standard parameters settings, which should only be adapted based on well-informed conceptual considerations.

**A. Experiment info file:**

| Article | n | x | y | z | Space | Domain |
|---|---|---|---|---|---|---|
| Study 1 | 13 | 2 | 14 | 41 | TAL | Cognitive Interference |
| Study 1 | 13 | -28 | -4 | 53 | TAL | Cognitive Interference |
| Study 1 | 13 | -37 | 23 | 25 | TAL | Cognitive Interference |
| Study 1 | 13 | 42 | 25 | 25 | TAL | Cognitive Interference |
| Study 1 | 13 | 56 | -49 | 27 | TAL | Cognitive Interference |
| Study 1 | 13 | -58 | -46 | 16 | TAL | Cognitive Interference |
| Study 1 | 13 | 39 | 47 | 9 | TAL | Cognitive Interference |
| Study 1 | 13 | 17 | -31 | 6 | TAL | Cognitive Interference |
| Study 1 | 13 | -57 | -51 | 6 | TAL | Cognitive Interference |
| Study 1 | 13 | -16 | -15 | 4 | TAL | Cognitive Interference |
| Study 1 | 13 | 50 | 24 | 3 | TAL | Cognitive Interference |
| Study 1 | 13 | -45 | 25 | -5 | TAL | Cognitive Interference |
| | | | | | | |
| Study 2 | 18 | -48 | 20 | 22 | MNI | Emotional Interference |
| Study 2 | 18 | -50 | 10 | 4 | MNI | Emotional Interference |
| Study 2 | 18 | 4 | 22 | 42 | MNI | Emotional Interference |
| Study 2 | 18 | -28 | -64 | 44 | MNI | Emotional Interference |
| Study 2 | 18 | -32 | -62 | -30 | MNI | Emotional Interference |
| Study 2 | 18 | 32 | -60 | -28 | MNI | Emotional Interference |
| Study 2 | 18 | 38 | -48 | -22 | MNI | Emotional Interference |
| Study 2 | 18 | 46 | -66 | -12 | MNI | Emotional Interference |
| Study 2 | 18 | 40 | -48 | 46 | MNI | Emotional Interference |
| Study 2 | 18 | 46 | 8 | 46 | MNI | Emotional Interference |
| Study 2 | 18 | -34 | -18 | 66 | MNI | Emotional Interference |
| Study 2 | 18 | -14 | 4 | 2 | MNI | Emotional Interference |
| | | | | | | |
| Study 3 | 9 | -16 | -70 | 56 | MNI | Emotional Interference |
| Study 3 | 9 | 40 | -16 | 40 | MNI | Emotional Interference |
| Study 3 | 9 | -34 | 8 | 34 | MNI | Emotional Interference |
| Study 3 | 9 | 12 | -84 | -44 | MNI | Emotional Interference |
| Study 3 | 9 | -26 | -52 | -22 | MNI | Emotional Interference |
| | | | | | | |
| Study 4 | 24 | 15 | 15 | 10 | TAL | Face |
| Study 4 | 24 | -26 | 21 | -11 | TAL | Face |
| Study 4 | 24 | 30 | 14 | -1 | TAL | Face |
| Study 4 | 24 | -4 | 42 | 37 | TAL | Face |
| Study 4 | 24 | 41 | 34 | 15 | TAL | Face |
| Study 4 | 24 | -52 | 23 | 27 | TAL | Face |

**B. Analysis info file:**

| | A | B | C | D | E | |
|---|---|---|---|---|---|---|
| 1 | M | CogInf | +Cognitive Interference | | | |
| 2 | M | EmoInf | +Emotional Interference | | | |
| 3 | C | CogInf | +Cognitive Interference | | | |
| 4 | | EmoInf | +Emotional Interference | | | |
| 5 | P21 | EmoInf_P21 | +Emotional Interference | | | |
| 6 | B21 | CogInf | +Cognitive Interference | | | |
| 7 | | EmoInf | +Emotional Interference | | | |
| 8 | Cluster | Clu_CogEmoFace | +Cognitive Interference | +Emotional Interferer | +Face | ? |
| 9 | | | | | | |

**C. Config file:**

config.yml

```
# Project Folder
project:
  analysis_info: "analysis_info.xlsx"
  experiment_info: "experiment_info.xlsx"


# ALE Parameters

parameters:

  # If enabled pools multiple experiments from same paper into one experiment;
  pool_experiments: False # Recommended only for "ALL" analyses

  # If enabled TFCE is in addition to standard thresholding methods calculated (Frahm et al., 2022)
  tfce_enabled: True

  # If enabled ALE map (and therefore all following maps) will be masked by ICBM 10% GM mask;
  gm_masking: True # Recommended: True

  # Size of bins used in MA histogram
  bin_steps: 0.0001 # Use and don't change: 0.0001

  # If enabled uses xgboost models to predict cutoffs instead of monte carlo simulation (Frahm et al., 2024);
  # If dataset>150, Foci of one/more experiment >150 and sample size of one/more experiment >300:
  # Please verify the results by calculating the analyses with the traditional permutation
  cutoff_predict_enabled: True # Recommended: True;


  # P-value required for significance
  # !! ONLY IF CUTOFF PREDICT IS DISABLED, OTHERWISE WON'T HAVE EFFECT !!
  significance_threshold: 0.05 # Recommended: 0.05

  # Preliminary cluster forming threshold used in cluster-level family wise error correction
```

Figure 1: Examples of the three files needed for calculating an ALE meta-analysis in JALE: (A) an experiment info file, containing the coordinates and experimental information to be meta-analyzed (B) an analysis info file defining the specific meta-analyses that should be calculated as well as (C) a configuration file defining input files and parameters.

## 2.1. Comparison to previously published meta-analytic results

We validated the results of JALE to results of previously published meta-analyses (Aziz-Safaie et al., 2024; Cieslik et al., 2023; Müller et al., 2017, Müller et al., 2025). To do so, we used the input data from the respective previous meta-analyses and calculated main effect and contrast analyses using JALE. We then calculated Pearson correlations and Jaccard index scores between the previous and JALE results. For most meta-analyses we found correlation coefficients and Jaccard Indices >0.95 (supplemental table S1), indicating high correspondence between toolboxes (see supplemental material for a detailed description of the results).

## 3. New implementations

JALE includes new options for significance thresholding encompassing the calculation of threshold-free cluster enhancement statistics (TFCE) and a rapid approach for multiple-comparisons correction based on machine-learning trained models. Furthermore, JALE offers novel analytical approaches including iterative probabilistic ALE (IPA) for robustness (sensitivity) analyses and for controlling differences in dataset sizes when investigating commonalities across different meta-analyses, balanced contrasts accounting for imbalanced datasets in meta-analytic contrasts as well as a clustering approach, providing the option to cluster neuroimaging experiments based on the similarity of their reported results.

### 3.1. Thresholding

#### *3.1.1. Cut-off prediction*

Based on our simulation studies (Eickhoff et al., 2016; Frahm et al., 2022; Frahm et al., 2023), a significance threshold correction for multiple comparisons is recommended to ensure that convergence is not driven by only one or two experiments. In ALE, these corrections normally rely on Monte-Carlo simulations, which, depending on the number of iterations and dataset size, can be quite time consuming. We recently (Frahm et al., 2025) developed an alternative approach of correcting for multiple comparison which uses machine-learning trained models to predict cFWE, vFWE and TFCE-FWE-corrected (see supplemental Figure S1 for an exemplary meta-analysis using the three different correction methods) significance thresholds. The prediction is based on experimental features (e.g. dataset size, number of foci, number of subjects), shows high prediction performance ($R^2 > 0.98$), and the model has been validated with several published meta-analytical datasets. As the use of cut-off prediction reduces the computation time of an ALE meta-analysis from hours to minutes, as compared to the traditional approach, we recommend to use cut-off prediction for all statistical correction methods when using JALE. Importantly, performance of cut-off prediction is only evaluated when the experimental characteristics of the input data correspond to those of the data on which the model was trained on, i.e. dataset size < 150, number of foci per experiment < 150, number of subjects per experiment < 300. JALE will print a warning when this is not the case, and we recommend verifying results by repeating the analysis with disabling cut-off prediction, using the original permutation procedure.

#### *3.1.2. Threshold-free cluster enhancement (TFCE)*

Beyond traditional cFWE and vFWE corrections for multiple comparisons, using TFCE statistics (Smith and Nichols, 2009) has become quite popular in neuroimaging analyses over the last decade. Based on large-scale simulations, Eickhoff et al. (2016) generally recommended using cFWE correction, which is based on comparing the empirical finding to a null distribution of maximum cluster sizes observed by chance. Recently, we additionally applied and evaluated TFCE (Frahm et al., 2022). While being computationally much more demanding, it performed comparable to cFWE correction. Given the high computational costs, we originally did not recommend using TFCE. However, for some situations the use of TFCE might be beneficial (Frahm et al., 2022). We assume that this is especially the case when small and more localized effects are expected, for example in subcortical regions where gray matter masking restricts or divides larger clusters. When cut-off prediction was introduced in 2025, the computational demands of TFCE have also been massively reduced, allowing for TFCE-based significance testing to be run in a reasonable amount of time. TFCE is therefore implemented in JALE, in addition to vFWE and cFWE corrections. Thus, TFCE-based results can now also be readily examined, which may be particularly interesting and useful for discovering more localized effects.

### 3.2. Robustness analyses and accounting for bias

Using a sufficient number of experiments and thresholding ALE results in an appropriate manner (e.g., including at least 17 to 20 experiments for cFWE-corrected thresholding of a meta-analysis across task-fMRI experiments) safeguards against the risk of meta-analytic results being driven by only one or two experiments (cf. Eickhoff et al., 2016). However, effects may sometimes still be sensitive to the inclusion or exclusion of specific (subsets of) experiments, calling for robustness analyses. Furthermore, when comparing meta-analytical results of two (or more) different datasets (e.g., cognitive vs. emotional interference), imbalanced dataset sizes result in a difference in statistical power. In the past, these issues have been handled through different approaches, such as stratified random subsampling (Langner et al., 2018), leave-one-out Jackknife analyses (e.g. Gu et al., 2019; Janouschek et al., 2021), or permutation-based subsampling procedures (e.g. Gu et al., 2019; Poudel et al., 2020; Zhang et al., 2021). While these approaches are not new, they were mostly ad-hoc analyses and have been conducted differently by different authors. Drawing from these ideas, in JALE, we have now implemented an advanced subsampling approach for running robustness analyses and handling imbalanced datasets in a standardized way, as will be detailed below.

#### *3.2.1. Iterative probabilistic ALE (IPA)*

By setting the type of analysis in the analysis-info file (first column) to ‘P’, an IPA analysis is conducted. For this, several subsamples are created by randomly leaving out a certain number of experiments (k) in each subsample/iteration (N-k). In the analysis-info file N-k is specified after ‘P’ (for example when setting the first column to ‘P21’, IPA with subsets of 21 experiments in each iteration will be calculated). Using N-1 will result in a leave-one-out analysis. Then an ALE is run on each subsample, and the resulting thresholded convergence maps of each iteration are binarized. In general, the number of iterations for subsampling (with replacement) is specified in the config-file, but fewer iterations will automatically be performed if the number of unique subsamples that can be created from the dataset is smaller (e.g., with N = 50 and k = 2, only 1225 unique subsamples can be drawn). Once all iterations are completed, the binarized maps across all subsampling analyses are averaged, revealing voxel-wise relative frequencies of significance across subsamples (see Fig.2 for a schematic illustration of the procedure). Larger values indicate a relatively high robustness of voxel-wise convergence, with a value of 1 indicating that significant convergence was found for the given voxel in all subsampling analyses. Voxels with low relative frequencies, in turn, reflect regions that show significant convergence only for some specific subsamples (i.e., showing little robustness overall). Importantly, the averaged binarized map across all subsamples is per default not masked with the main effect observed in the full sample.

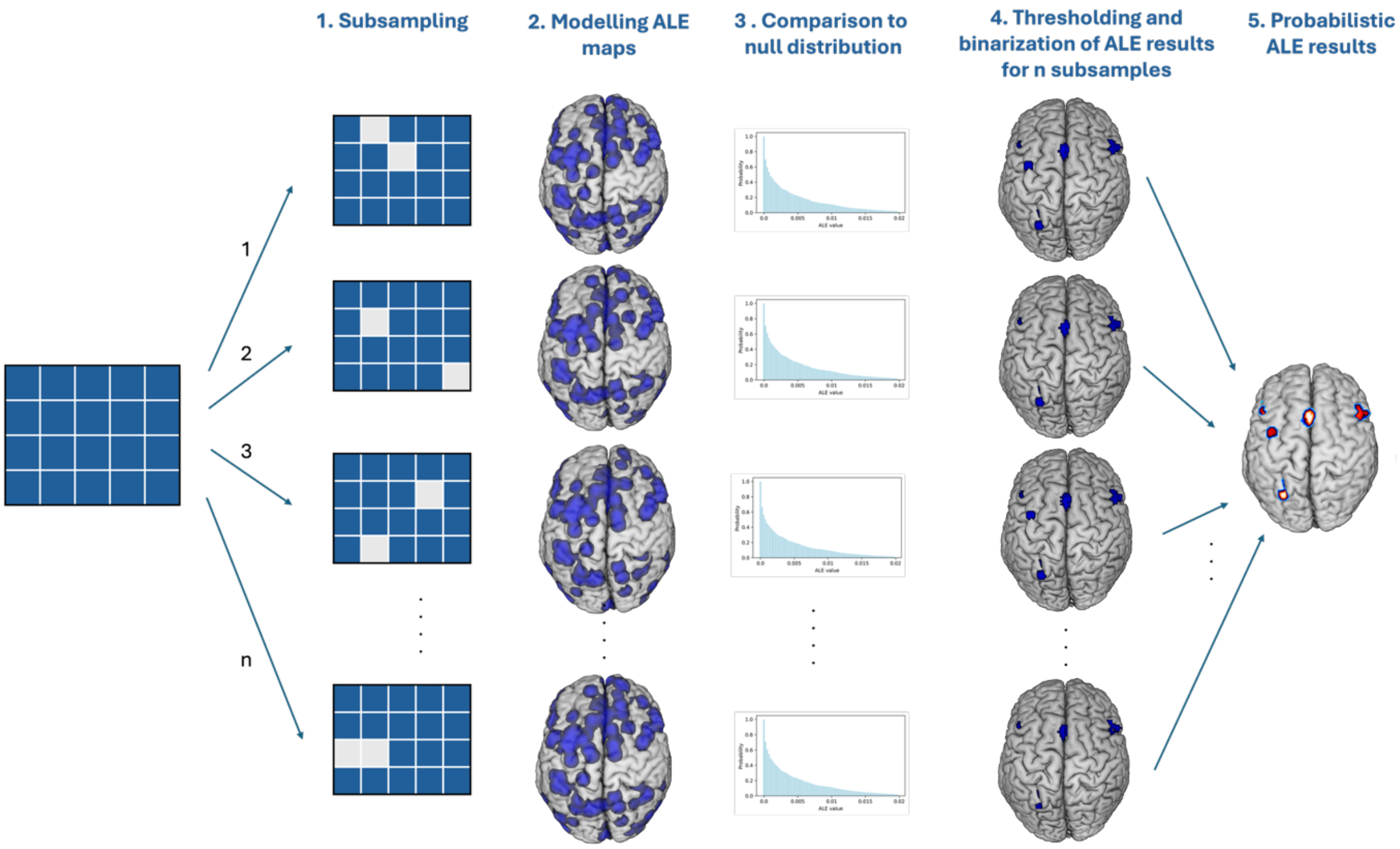

Figure 2: Schematic illustration of the Iterative Probabilistic ALE (IPA) procedure

On the one hand, IPA can be used to quantify the robustness of meta-analytic main effects; on the other hand, it can serve to accommodate for analyses that are strongly overpowered. For the former purpose, IPA is calculated by leaving out a rather small number of experiments (e.g., 10% of the total sample). For illustration, we conducted IPA for a dataset of 23 experiments investigating emotional interference. Figure 3 shows the main effect (A) as well as IPA results (B) based on randomly subsampling 90% of the experiments (i.e., n = 21; implemented by setting the type of analysis to ‘P21’ to define the subsample size as N−2) in each iteration. IPA results show high frequencies for all clusters of the main effect, corroborating their stability. Additionally, IPA revealed a cluster within left posterior middle temporal gyrus, which, however, showed a very low average frequency (i.e., little robustness). As this cluster did not reach significance in the standard main effect, the IPA results support the notion that conducting ALE meta-analyses with the recommended minimum number of experiments and thresholding approaches reveals stable results. If the research interest focuses on significant clusters of the main-effect analysis across the full dataset, IPA results should be inclusively masked by the respective thresholded main effect. Furthermore, when reporting IPA results, the number of experiments left out per sample and the number of iterations should always be indicated.

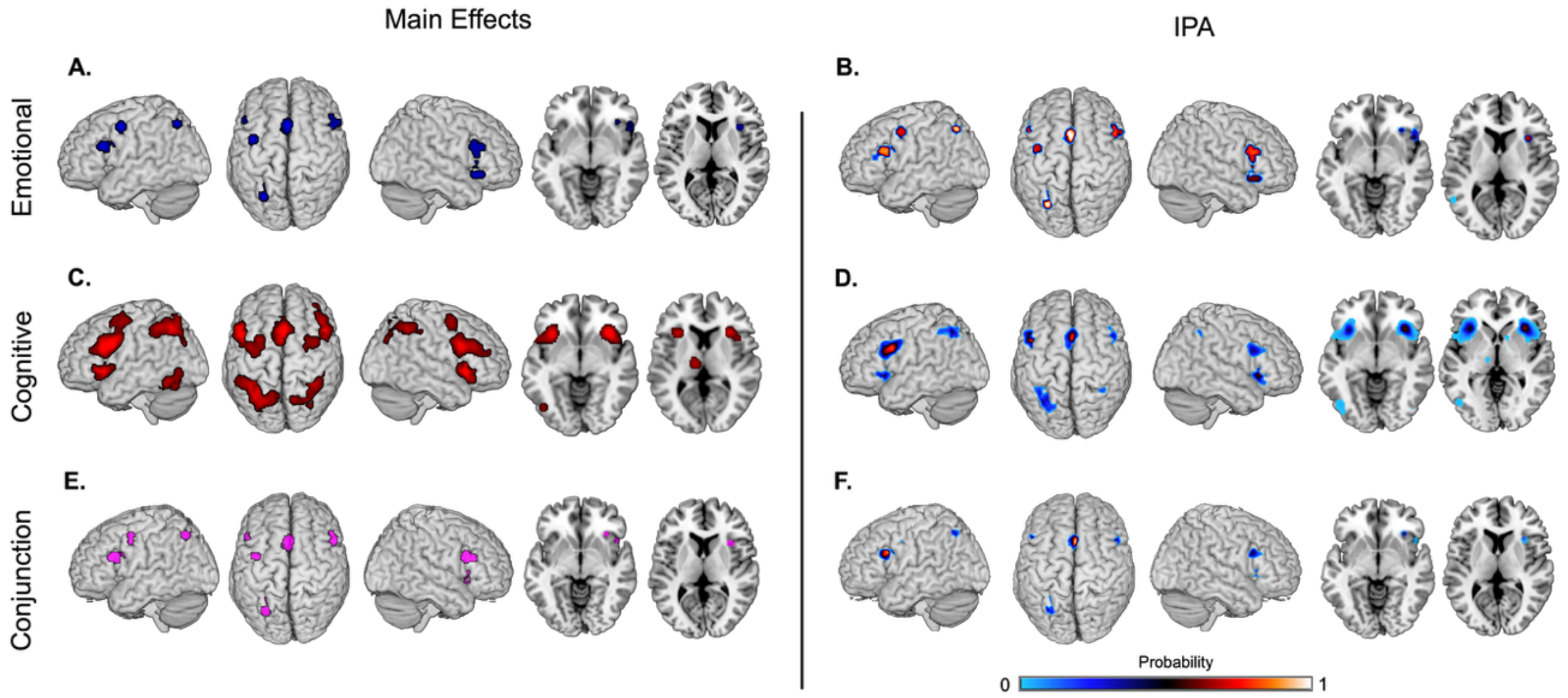

Figure 3: Results of standard ALE versus iterative probabilistic ALE (IPA) meta-analyses of neuroimaging experiments probing emotional or cognitive interference, respectively. (A) Full-sample main effect across 23 experiments. (B) IPA based on randomly selecting 21 emotion interference experiments (leave 10% out) per iteration. (C) Full-sample main effect across 154 cognitive interference experiments. (D) IPA based on randomly selecting 21 cognitive interference experiments per iteration. (E) Minimum conjunction across the standard main effects of emotion and cognitive interference. (F) Minimum conjunction across the IPA results of emotion and cognitive interference.

IPA can additionally be used to compare the stability or robustness of effects between a largely overpowered dataset and a relatively small one. In this case, comparability is reduced because the large dataset allows for detecting relatively small effects being present only in a subset of studies, while this is almost impossible in the smaller dataset. By performing IPA on the same number of randomly chosen experiments (e.g., 21 experiments per iteration) for each dataset, this approach can account for the higher power in the larger dataset. To illustrate this, we used the same small example dataset of emotional interference experiments and, additionally, a large dataset of cognitive interference experiments (N = 154). Figure 3A and 3C shows the main effect for either dataset, with a much broader brain network consistently involved in processing cognitive interference. Running IPA with subsampling 21 experiments from either dataset revealed that some clusters of convergence observed for the main effect of cognitive interference could only be found in some subsamples (e.g., left thalamus and left fusiform gyrus; Fig 3D), indicating decreased robustness. This may be due to increased heterogeneity in the larger dataset, with some effects being present only in a subset of cognitive interference studies. Importantly, this does not indicate that these clusters are not relevant or should not be interpreted, it just indicates that similarly small effects may not be detected in the smaller dataset due to its lower overall power. Adjusting the power of the larger dataset thus makes robustness analyses more comparable.

IPA results may also be useful for the investigation of similarities between two meta-analytic results, i.e. for calculating conjunctions. This allows moving from the standard binary viewpoint asking ‘Which voxels are found in both analyses?’ to a continuous perspective of combined probabilities asking ‘How probable is a given voxel’s consistent involvement in both analyses?’. Figure 3E and 3F illustrates the conjunctions between emotion and cognitive interference across the standard main effects (3E) and IPA results (3F), respectively.

#### *3.2.2. Balanced contrasts*

In the context of meta-analytic contrast/subtraction analyses, a strong imbalance in the sample size between two datasets, like in our example of cognitive and emotional interference, might create a relevant bias in detecting significant group differences. It is assumed that the label-permutation procedure of the standard ALE meta-analytic contrast analysis addresses this bias by preserving the imbalance also in the null distribution. That is, the difference in ALE values between group A and group B are compared to a null distribution of differences when experiments are reassigned between groups while preserving the original group sizes. Nevertheless, based on our experience, reviewers often raise concerns regarding contrast analyses with strongly unequal sample sizes.

In a similar vein as for conjunctions, difference maps between IPA results can be calculated revealing a voxel-wise map of difference probability scores indicating for which of the two datasets the probability of convergence is higher when the number of included experiments is held constant across the two datasets. To allow for statistical inference, we also implemented a subsampling approach for the subtraction analyses, i.e. balanced contrasts, which also compares resulting differences to a null distribution of difference scores and corrects for multiple comparisons. In a first step, multiple same-sized subsamples are sampled from group A and B in a way that matches the number of experiments between them. In a next step, ALE maps are computed for these subsamples, differences between A and B are calculated per iteration, and then these difference maps are averaged across all iterations. This average difference is then compared to a null distribution of differences. The null is created by first generating, for each of the two groups, several random datasets of the same size as the original ones, where the foci are randomly distributed across all grey-matter voxels. From each of these random datasets, again multiple same-sized subsamples are drawn, ALE maps computed, difference maps calculated per iteration and then averaged across iterations. For a schematic overview of this procedure, see Figure 4.

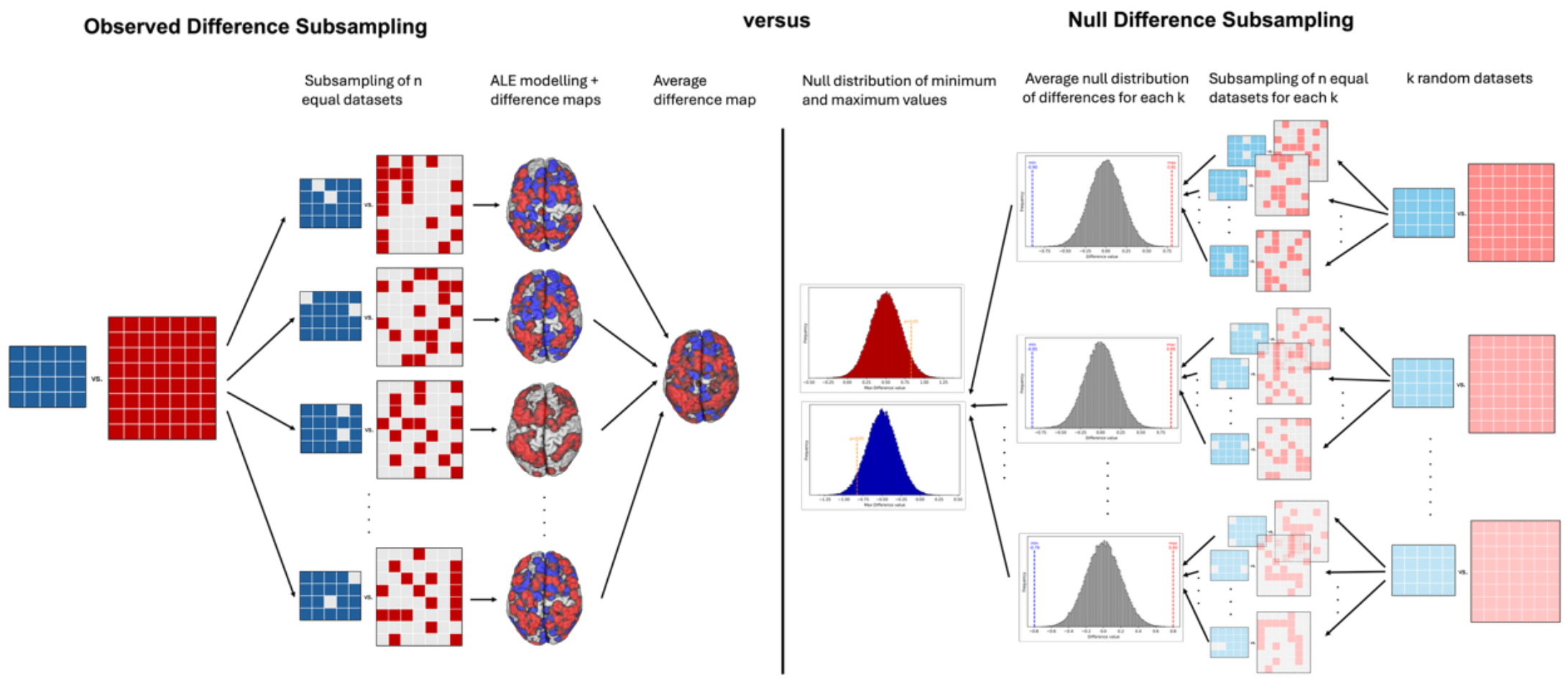

Figure 4: Schematic illustration of the procedure for calculating balanced contrasts.

Due to the repeated subsampling (random datasets x subsample n), without using a machine-learning trained model for predicting significance thresholds of differences, the approach of balanced contrasts is computationally highly expensive. Also, the approach needs further systematic evaluation regarding its ability to account for potential bias from dataset size imbalances. Therefore, we currently do not recommend balanced contrast as the standard approach for contrast analyses but rather as a tool to explore the sensitivity of ALE group differences to sample size differences.

### 3.3. Clustering

Beyond traditional neuroimaging meta-analyses that test for spatial convergence across experiments that are grouped based on prior knowledge (e.g., for answering questions like 'Where does brain activation converge across studies investigating cognitive interference?'), there is also the possibility of a data-driven approach, grouping experiments based on the similarity of their reported coordinates (e.g., answering questions like 'Are there any distinct subgroups of experiments investigating cognitive interference based on the brain regions they involve?'). This approach has previously been used to cluster neuroimaging studies of face (Laird et al., 2015), reward (Flannery et al., 2020), or emotion processing (Morawetz et al., 2020; Riedel et al., 2018), social cognition (Schurz et al., 2021), naturalistic perception paradigms (Bottenhorn et al., 2019), as well as experiments in psychiatric diseases (Boisvert et al., 2024; Dugré et al., 2022).

In JALE, the clustering of experiments based on their reported coordinates can be examined by setting the type of analysis in the analysis-file (first column) to 'Cluster'. The input is the same as for the classic meta-analysis, and clustering from k=2 to k=n is computed as specified in the config-file.

The implementation is similar as in Dugré et al. (2022) and Boisvert et al. (2024): In a first step, for each experiment, a modelled activation map is computed in the same way as for standard ALE meta-analysis. Then, each of these maps is converted into a 1-D vector of modelled activation values for all grey-matter voxels, and then the vectors are correlated (Spearman correlation or Pearson as defined in the config-file) between every experiment and every other experiment. K-medoids or agglomerative hierarchical clustering is then performed based on the correlation distance (1-ρ, or 1-r) between experiments. For hierarchical clustering, there is the choice between using complete or average linkage.

For robustness, the clustering is not only done once on the full dataset; rather, a subsampling procedure without replacement is employed. Consistent cluster labels across iterations are identified for each clustering solution (k) by first calculating for each experiment pair the Hamming distance for each iteration (considering only the experiments that are selected for the respective iteration). Second, using the same clustering method (k-medoids or hierarchical clustering with the respective linkage method), clustering is performed on the averaged Hamming distances between experiments.

Several commonly used metrics for evaluating cluster solutions are calculated: Silhouette coefficient (SI), Calinski-Harabasz index (CH), and separation density for hierarchical clustering; and SI, CH, and variation of information for k-medoids clustering. SI and CH indices assess how similar each experiment is to the other experiments in its own cluster (i.e., within-cluster similarity / compactness) relative to how far it is from experiments in the other clusters (between-cluster distance). Separation-Density is quantified as the ratio of the number of experiments of the larger child cluster to the number of experiments of the original (parent) cluster after a hierarchical split. Variation of Information (VI) assesses the amount of information that changes when moving from a solution with k-1 clusters to a solution with k clusters. Each metric is calculated for each iteration and averaged across iterations. SI and CH are additionally normalized to a null distribution of clustering of random modelled activation maps. To do so, several random datasets of the same size as the original datasets are created by shuffling the coordinates across experiments but preserving the properties of the experiments (number of foci, sample size) of the true dataset. For each random dataset, clustering is performed, and SI and CH values are calculated. A z-value is then computed for each k and metric by subtracting the average metric of the null distribution (SI or CH, respectively) from the true average metric divided by the standard deviation (standard deviation of the null or pooled standard deviation as defined in the config-file).

Figure 5 shows the results of an example of a (k-medoid) clustering analysis where we combined the experiments of cognitive and emotion interference (187 experiments) and added coordinates of experiments of an additional dataset of emotional face processing (56 experiments). SI and CH scores both point to k=2 as the preferred clustering solution (Figure 5A), while comparison to a null distribution of clustering of random modelled activation maps showed that all solutions reveal higher SI and CH values than what would be expected by chance (Figure 5B). To investigate the networks driving the separation, conventional ALE meta-analyses can be calculated separately across experiments that are clustered into different meta-analytical groupings (MAGs). For our example dataset, using the labels of k=2, ALE meta-analyses revealed for MAG-1 (Figure 5C in green) convergence in a broad network that strongly involves regions of the so-called (extended) multiple-demand network (Duncan, 2010; Camilleri et al., 2019), while for MAG-2 (Figure 5C in purple), some of those regions are also found but with an additional consistent involvement of the amygdalae, fusiform and inferior occipital gyrus as well as right posterior superior temporal gyrus.

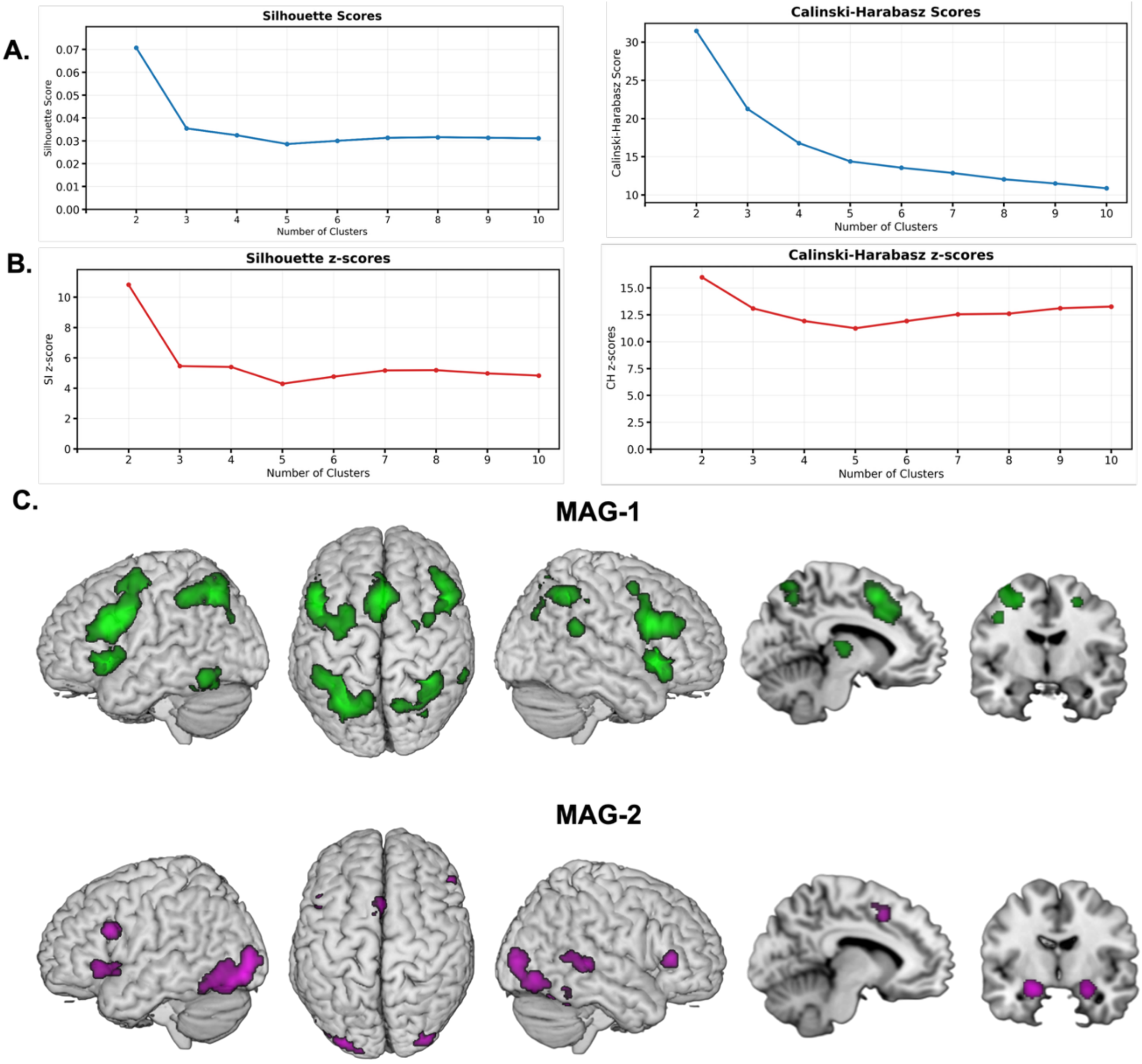

Figure 5: Example of a clustering analyses of experiments probing interference and emotional face processing. (A) Silhouette (SI) and Calinski-Harabasz (CH) scores for different k's. (B) z-scores for different k's comparing SI and CH against a null distribution of clustering random modelled activation maps. (C) Results of ALE meta-analyses across experiments grouped into the same meta-analytic grouping (MAG) for the k=2 clustering solution.

Summarizing, meta-analytic clustering may be used as a complementary approach to standard top-down meta-analyses as it groups experiments based on the similarity of activation maps modeled from reported brain coordinates. This can then help to reveal latent data structure not captured by the standard way of organizing experiments in a top-down, knowledge-driven way.

## 4. Conclusion

We here present JALE, a new, freely available, easy-to-use Python toolbox for ALE meta-analysis. This package brings together established ALE main effect and contrast analyses with recently developed approaches for significance thresholding as well as analyses that test for robustness, account for biases in dataset sizes, and cluster meta-analytical input data. We invite the neuroimaging community to explore the new toolbox, test its various possibilities and use it for answering new scientific questions.

## Statements and Declarations

### Declaration of the use of AI

DeepL write (https://www.deepl.com/en/write) was used for editing existing text passages for precision, formality, and clarity. The Wiki-documentations and docstrings in the code of JALE were written with the support of ChatGPT (version 3.5).

### Data and Code Availability

The JALE toolbox and user instructions can be found on github: https://github.com/inm7/jale

### Declaration of Competing Interests

The authors declare no potential conflict of interest.

### Supplementary Material

Supplementary Material is provided

**Supplemental Material**

**Comparison to previously published meta-analyses**

We validated the results based on the new JALE toolbox to results of published previous meta-analyses (Aziz-Safaie et al., 2024; Cieslik et al., 2023; Müller et al., 2017, Müller et al., 2025) by calculating Pearson correlations and Jaccard index scores between the previous results and those using the JALE toolbox. For the main-effects we used unthresholded z-maps, ALE maps and thresholded cFWE corrected maps (except for Müller et al. (2017) where there were no significant cFWE results), for the conjunction analysis we used the minimum conjunction maps and for the contrast analyses the thresholded z-maps were utilized. For most meta-analyses we found correlation coefficients and Jaccard Indices >0.95 (supplemental table S1), indicating high correspondence between toolboxes. Lower correspondence was found for contrast analyses, mainly due to small differences in cluster-sizes. For one analysis, i.e. Aziz-Safaie et al. (2024), for the main effect of the standard Go/NoGo task and the minimum conjunction involving this main effect relatively low values for the comparison of cFWE maps were found (while ALE maps and unthresholded z-maps corresponded well). This is most likely due to differences in precision of the arithmetic float in the Matlab and Python version of the toolboxes, leading to slight differences in z-values. In most of the cases this didn't have an effect, but for some rare "at threshold cases", like we observed for the main effect of the standard Go/NoGo task, this can then lead to discrepancies between both toolboxes.

Table S1. Comparison between previously published results and those using the new JALE toolbox

| **Analysis** | **ALE map Pearson r / Jaccard Index** | **z- map Pearson r / Jaccard Index** | **Thresholded map Pearson r / Jaccard Index** |
|---|---|---|---|
| ***Aziz-Safaie et al. (2024)*** | | | |
| Main effect Go/NoGo | 1/1 | 1/1 | 0.995 / 0.986 |
| Main effect standard Go/NoGo | 1/1 | 1/1 | 0.943 / 0.874 |
| Main effect complex Go/NoGo | 1/1 | 1/1 | 0.993 / 0.979 |
| Main effect Stop Signal | 1/1 | 1/1 | 0.998 / 0.993 |
| Main effect standard Stop Signal | 1/1 | 1/1 | 1/1 |
| Main effect complex Stop Signal | 1/1 | 1/1 | 0.993 / 0.979 |
| Minimum conjunction across standard Go/NoGo and standard Stop Signal | - | - | 0.656 / 0.478 |
| Minimum conjunction across complex Go/NoGo and complex Stop Signal | - | - | 0.994 / 0.982 |
| ***Cieslik et al., 2023*** | | | |
| Main effect commission error | 1/1 | 1/1 | 1 / 1 |
| Main effect correct interference resolution | 1/1 | 1/1 | 0.999 / 0.994 |
| Contrast between error and correct | - | - | 0.934 / 0.982 |
| Minimum conjunction across error and correct | - | - | 0.995 / 0.983 |
| ***Müller et al. 2017*** | | | |
| Overall analysis | 1/1 | 1/1 | - |
| Main effect aberrant emotion processing | 1/1 | 1/1 | - |
| Main effect aberrant cognitive processing | 1/1 | 1/1 | - |
| ***Müller et al., 2025*** | | | |
| Main effect Incongruent vs. Congruent | 1/1 | 1/1 | 1 / 1 |
| Main effect Incongruent vs. Neutral | 1/1 | 1/1 | 1 / 1 |
| Contrast between IvsC and IvsN | - | - | 0.985 / 0.939 |
| Minimum conjunction across IvsC and IvsN | - | - | 1/ 1 |
| Main effect blocked design | 1/1 | 1/1 | 0.996 / 0.986 |
| Main effect mixed design | 1/1 | 1/1 | 1 / 1 |
| Contrast between blocked and mixed | - | - | 0.965 / 0.97 |
| Minimum conjunction between blocked and mixed | - | - | 0.997 / 0.99 |
| Main effect additional cognitive demand | 1/1 | 1/1 | 0.999 / 0.995 |
| Main effect no additional cognitive demand | 1/1 | 1/1 | 0.998 / 0.991 |
| Contrast between additional demand and no demand | - | - | 0.969 / 0.887 |
| Minimum conjunction across demand and no demand | - | - | 1 / 0.999 |
| Main effect colorword | 1/1 | 1/1 | 1 / 1 |
| Main effect emotional | 1/1 | 1/1 | 1 /1 |
| Main effect other | 1/1 | 1/1 | 0.991 / 0.973 |
| Contrast between colorword and emotional | - | - | 0.971 / 0.93 |
| Minimum conjunction across colorword and emotional | - | - | 1/1 |
| Contrast between colorword and other | - | - | 0.982 / 0.944 |
| Minimum conjunction across colorword and other | - | - | 0.993 / 0.982 |
| Contrast between emotional and other | - | - | 0.983 / 0.773 |
| Minimum conjunction across emotional and other | - | - | 0.99 / 0.972 |

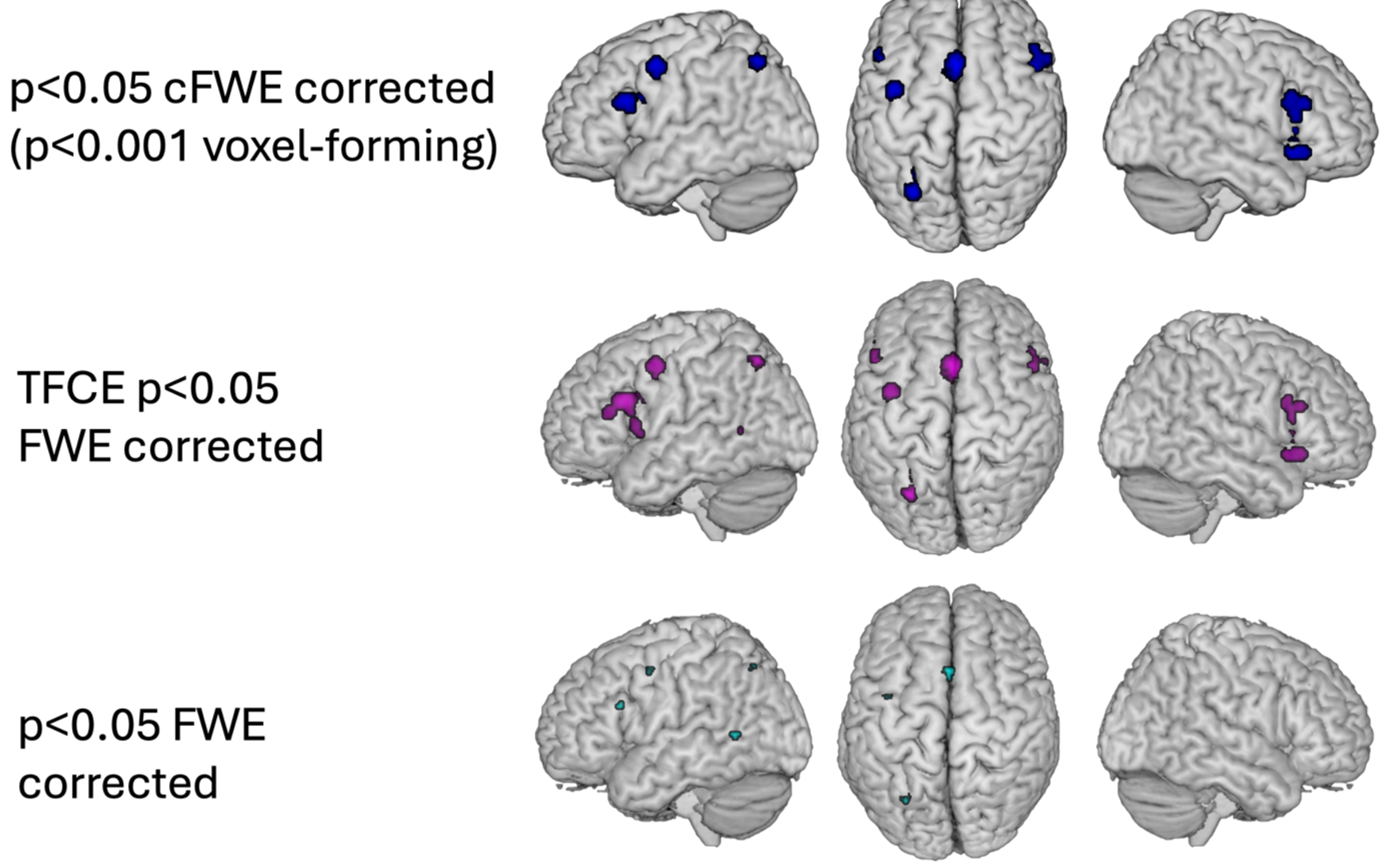

Figure S1. Illustration of the results of different thresholding options using cut-off prediction in JALE based on a meta-analysis of 23 experiments of emotional interference.